\documentclass[12pt]{article}
\usepackage[cp1251]{inputenc}
\usepackage{graphicx}
\usepackage{amssymb}

\fontsize{14}{24pt}\selectfont

\title{A note on scattering in deformed space with minimal length}
\author{M. M. Stetsko\footnote{E-mail: mykola@ktf.franko.lviv.ua}\ \ 
\\
  {\small Department for Theoretical Physics, Ivan Franko National University of Lviv,}\\
{\small 12 Drahomanov St., Lviv, UA-79005, Ukraine}}
\begin{document}
\maketitle

\abstract{We consider the elastic scattering in deformed space
with minimal length. We give the basic relation for the elastic
scattering in deformed space. We also investigate the partial wave
method in deformed space. It is shown that the relations for the
scattering amplitude and cross-section formally coincides with
ordinary ones.}

\section{Introduction}
In the recent years, a lot of works have been devoted to the
quantum mechanical problems in a deformed space with minimal
length. Such an interest was motivated by several independent
lines of investigation as the string theory \cite{gross} and
quantum gravity \cite{maggiore} where the existence of a finite
lower bound for the possible resolution of length \cite{witten}
was proposed. Kempf and collaborators showed that finite
resolution of length can be obtained from the deformed commutation
relations \cite{kempf1,kempf2,kempf3,kempf4}. One should note that
deformed commutation relations were introduced earlier by Snyder
\cite{snyder}.

The deformed algebra leading to the existence of a minimal length
in a $D$-dimensional case takes form
\begin{eqnarray}\label{algebra}
\begin{array}{l}
[X_i, P_j]=i\hbar(\delta_{ij}(1+\beta P^2)+\beta'P_iP_j), \quad
[P_i, P_j]=0,
\\
\\
\displaystyle {[X_i,
X_j]}=i\hbar\frac{(2\beta-\beta')+(2\beta+\beta')\beta
P^2}{1+\beta P^2}(P_iX_j-P_jX_i),
\end{array}
\end{eqnarray}
where $\beta$, $\beta'$ are the deformation parameters. We assume
that these quantities are nonnegative $\beta,\beta'\geqslant0$. It
is easy to show that minimal length equals
$\hbar\sqrt{\beta+\beta'}$.

Deformed Heisenberg algebra (\ref{algebra}) causes new
complications in solving quantum mechanical problems. It is well
known that algebra (\ref{algebra}) does not have the position
representation \cite{kempf2}. There are just a few known problems
for which the energy spectra have been found exactly
\cite{Tkachuk1,Tkachuk2,chang,Dadic,quesne,fityo}.

The several works were devoted to the hydrogen atom in the space
with deformed Heisenberg algebra (\ref{algebra}). In work
\cite{Brau} the partial case of deformation was considered when
$2\beta=\beta'$, the spectrum was found perturbatively. In paper
\cite{Benczik} a general case $2\beta\neq \beta'$ was considered,
using perturbation theory the spectrum of the hydrogen atom was
found but when $l\neq 0$. For the $s$-states the corrections to
the energy levels were found numerically. In work \cite{mykola} a
modified perturbation theory was proposed that allowed to obtain
the analytical correction to the $s$-levels. In paper
\cite{stetsko2} the corrections to the energy for arbitrary
$s$-levels of the hydrogen atom were calculated

The scattering problem in deformed space was investigated in
\cite{st_tk_2}. The elastic scattering was considered. Here we
carry on the investigation of scattering problem. We develop the
partial wave method and compare it with previous results.

\section{Scattering amplitude and differential cross-section}
To investigate the scattering problem it is essential to introduce
the representation of operators satisfying the algebra
(\ref{algebra}). It is well known that such an algebra has the
momentum representation, but we use the approximate representation
fulfilling the algebra in the first order over the deformation
parameters \cite{mykola,st_tk_2}:
\begin{eqnarray}\label{rep1}
\left\{
\begin{array}{l}
\displaystyle
X_i=x_i+\frac{2\beta-\beta'}{4}\left(x_ip^2+p^2x_i\right),
\\
\displaystyle P_i=p_i+\frac{\beta'}{2}p_ip^2,
\end{array}
\right.
\end{eqnarray}
where $p^2=\sum^3_{j=1}p^2_j$ and the operators $x_i$, $p_j$
satisfy the canonical commutation relation. The position
representation $x_i=x_i$, $p_j=i\hbar\frac{\partial}{\partial
x_j}$ can be taken for the ordinary Heisenberg algebra.  We note
that in the special case $2\beta=\beta'$ the position operators
commute in linear approximation over the deformation parameters,
i.e. $[X_i,X_j]=0$.

The Schr\"{o}dinger equation with arbitrary potential $U({\bf R})$
in canonical variables takes the following form:
\begin{equation}\label{Schroedinger1}
\left(\frac{p^2}{2m}+\frac{\beta'p^4}{2m}+U(\textbf{r},\textbf{p})\right)\Psi=E\Psi.
\end{equation}
We suppose that $U({\bf r},{\bf p})\rightarrow 0$ when
$r\rightarrow \infty$ and motion of a scattered particle at large
distances form the scattering center is free. The kinetic energy
of a free particle reads:
\begin{equation}\label{kinetic_energy}
E=\frac{\hbar^2k^2}{2m}(1+\beta'\hbar^2k^2),
\end{equation}
where ${\bf k}$ is the wave vector of an incident particle and
${\bf P}=\hbar{\bf k}(1+\beta'\hbar^2k^2/2)$ is the momentum of
the particle. We investigate the elastic scattering so we have
that after the scattering $k=k'$, where ${\bf k}'$ is the wave
vector of scattered particle.

Then we can write a formal solution of equation
(\ref{Schroedinger1}) in the form:
\begin{equation}\label{solution}
\Psi(\textbf{r})=\psi_k(\textbf{r})+\int
G(\textbf{r},\textbf{r}')\frac{2m}{\hbar^2}U(\textbf{r}',\textbf{p}')\Psi(\textbf{r}')\textrm{d}\textbf{r}',
\end{equation}
where $G({\bf r},{\bf r}')$ is the Green's function.

As was shown in paper \cite{st_tk_2} the asymptotic Green's
function takes form:
\begin{equation}\label{GreenSimplified}
G(|{\bf r}-{\bf r}'|)=-\frac{1}{4\pi|{\bf r}-{\bf
r}'|(1+2\beta'\hbar^2k^2)} e^{ik|{\bf r-r'}|}.
\end{equation}

Using this Green's function (\ref{GreenSimplified}) and after some
simplifications we rewrite equation (\ref{solution}) in the form:
\begin{equation}\label{wavefunction}
\Psi(\textbf{r})=\psi_k(\textbf{r})-\frac{m}{2\pi\hbar^2(1+2\beta'\hbar^2k^2)}\frac{e^{ikr}}{r}\int
e^{-i\textbf{k}'\textbf{r}'}U(\textbf{r}',\textbf{p}')\Psi(\textbf{r}')\textrm{d}\textbf{r}',
\end{equation}
where ${\bf k}'=k{\bf n}$.

Equation (\ref{wavefunction}) can be represented in the form:
\begin{equation}\label{WaveFunct2}
\Psi(\textbf{r})=e^{i{\bf kr}}+\frac{e^{ikr}}{r}f,
\end{equation}
where
\begin{equation}\label{scattering_amplitude}
f=-\frac{m}{2\pi\hbar^2(1+2\beta'\hbar^2k^2)}\int e^{-i{\bf k'r'
}}U(\textbf{r}',\textbf{p}')\Psi(\textbf{r}')\textrm{d}\textbf{r}'
\end{equation}
is the scattering amplitude.

As we see the first term in equation (\ref{WaveFunct2})
corresponds to the wave function of an incident particle and the
second term corresponds to the wave function of a scattered
particle.

It is well known that central problem of the scattering theory is
the calculation of the differential cross-section. It was shown in
\cite{st_tk_2} that differential cross-section in deformed space
takes the same form as in the ordinary quantum mechanics:
\begin{equation}\label{cross-section_Coulomb}
\frac{d\sigma}{d\Omega}=|f|^2.
\end{equation}

The last equation allows to calculate the differential
cross-sections for an arbitrary potential of scattering. But as we
can see from expression (\ref{scattering_amplitude}) the
scattering amplitude and as a consequence differential
cross-section can not be calculated exactly. It necessary to use
the approximate calculation. In the Born approximation we consider
the scattering potential as a small perturbation and solve
equation (\ref{wavefunction}) by the method of successive
approximation. In the first approximation we substitute the plane
wave $\psi_{{\bf k}}({\bf r})=\exp{(i{\bf kr})}$ in relation
(\ref{scattering_amplitude}).

As was shown in \cite{st_tk_2} the differential cross-section for
the Coulomb potential $U=e^2/R$, where $R=\sqrt{\sum_iX_i^2}$ in
deformed space takes the form:
\begin{eqnarray}\label{diff_cross-sect_Coul}
\begin{array}{c}
\displaystyle\frac{d\sigma}{d\Omega}=\frac{m^2e^4}{4\hbar^4k^4\sin^4\frac{\vartheta}{2}}+
\frac{me^2}{\hbar^2k^2\sin^2\frac{\vartheta}{2}}\left(\frac{me^2}{2}(2\beta-\beta')\times\right.
\\
\\
\displaystyle\left.\left[\ln\left(\hbar^2(2\beta-\beta')k^2\sin^2\frac{\vartheta}{2}\right)+2\gamma-1-
\frac{1}{2\sin^2\frac{\vartheta}{2}}\right]-\beta'\frac{me^2}{\sin^2\frac{\vartheta}{2}}\right).
\end{array}
\end{eqnarray}

The first term in (\ref{diff_cross-sect_Coul}) is the ordinary
differential cross-section and other terms are caused by
deformation of commutation relations. As we see the corrections to
the differential cross-section nonanalytically depend on the
deformation parameters, but when the deformation parameters tend
to zero the corrections also tend to zero and we return to the
ordinary differential cross-section. We also note that as in
ordinary quantum mechanics the scattering amplitude for the
particle in the Coulomb potential is ill defined in the Born
approximation, so it is necessary to calculate the scattering
amplitude for the Yukawa potential $U(R)=-e^2\exp{(\lambda R)}/R$
and then one should tend the parameter $\lambda$ to zero.

\section{Partial wave method}
When the potential of scattering is spherically symmetric then the
angular momentum of scattering particle is the integral of motion,
so it should to note that particles with different orbital quantum
numbers are scattered independently. Then as in ordinary quantum
mechanics we can represent the scattering cross-section as a sum
of partial cross-sections for the fixed values of the orbital
quantum numbers $l$. So we assume that we have a spherically
symmetric scattering potential, and consider the scattering
process on this potential. Arbitrary solution of the
Shr\"{o}dinger equation can be represented in the form:
\begin{equation}\label{decomp_part_wave}
\Psi({\bf r})=\sum_{l=0}^{+\infty} A_lR_{kl}(r)P_l(\cos\vartheta),
\end{equation}
where $R_{nl}(r)$ are the radial wave functions and
$P_l(\cos{\vartheta})$ are the Legendre polynomials.

Firstly consider free particles. The Shr\"{o}dinger equation for a
free particle:
\begin{equation}
-\frac{\hbar^2\nabla^2}{2m}(1-\beta'\hbar^2\nabla^2)\psi
=\frac{\hbar^2k^2}{2m}(1+\beta'\hbar^2k^2)\psi,
\end{equation}
or in equivalent form:
\begin{equation}\label{factorized_equation}
(\nabla^2+k^2)(1-\beta'\hbar^2(\nabla^2-k^2))\psi=0.
\end{equation}
We represent the solution of last equation as sum of two functions
$\psi_1$ and $\psi_2$ which are the solutions of the following
equations:
\begin{equation}\label{factorized_1}
(\nabla^2+k^2)\psi_1=0,
\end{equation}
\begin{equation}\label{factorized_2}
(1-\beta'\hbar^2(\nabla^2-k^2))\psi_2=0.
\end{equation}
Radial part of wave function $\psi_1$ takes the well known form:
\begin{equation}\label{rad_free_particle_norm}
R_{kl}(r)=\frac{1}{k^l}\sqrt{\frac{2}{\pi}}r^l\left(-\frac{1}{r}\frac{d}{dr}\right)^l
\frac{\sin{kr}}{r}.
\end{equation}
Equation (\ref{factorized_2}) gives us unphysical solution and we
reject it.

The asymptotic of solution (\ref{rad_free_particle_norm}) at large
distances reads:
\begin{equation}\label{asympt_free}
R_{kl}(r)=\sqrt{\frac{2}{\pi}}\frac{\sin{\left(kr-l\frac{\pi}{2}\right)}}{r}.
\end{equation}

We suppose similarly as in ordinary quantum mechanics that
asymptotic of the wave function for the scattered particle is the
same and differs from (\ref{asympt_free}) by the phase shift only:
\begin{equation}\label{asymp_scatt}
\overline
R_{kl}(r)=\sqrt{\frac{2}{\pi}}\frac{\sin{\left(kr-l\frac{\pi}{2}+\delta_l\right)}}{r},
\quad r\rightarrow\infty.
\end{equation}

We use decomposition (\ref{decomp_part_wave}) and instead of
function $R_{kl}(r)$ we take wave function (\ref{asympt_free}) for
the free particle and function (\ref{asymp_scatt}) for the
particle in the external field. Then we substitute aforementioned
decompositions in (\ref{WaveFunct2}), so we obtain:
\begin{eqnarray}
\psi_{scatt}=f\frac{e^{ikr}}{r}=\frac{1}{kr}\sum^{+\infty}_{l=0}i^l(2l+1)
P_l(\cos{\vartheta})\times \nonumber
\\
\left[C_l\sin\left(kr-l\frac{\pi}{2}+\delta_l\right)
-\sin\left(kr-l\frac{\pi}{2}\right)\right].
\end{eqnarray}
After simple transformations we obtain:
\begin{equation}\label{scatt_amplitude_partial}
f=\frac{i}{2k}\sum^{+\infty}_{l=0}(2l+1)(1-e^{2i\delta_l})P_l(\cos{\vartheta}).
\end{equation}
Last relation formally coincides with the expression for the
scattering amplitude in the ordinary quantum mechanics, but we
note that wave number $k$ and phase shift $\delta_l$ depend on the
deformation parameters.

Using relation (\ref{cross-section_Coulomb}) and after integration
over all spatial angles we have:
\begin{equation}
\sigma=\frac{\pi}{k^2}\sum^{+\infty}_{l=0}(2l+1)\sin^2{\delta_l}.
\end{equation}
Similarly as in ordinary quantum mechanics the scattering
cross-section relates to the imaginary part of the scattering
amplitude in forward direction, so we have:
\begin{equation}
\sigma=\frac{4\pi}{k}{\rm Im}f(0).
\end{equation}
Partial wave method also allows to obtain the phase shifts. Using
the radial functions $\chi_{kl}(r)=rR_{kl}(r)$ and
$\overline\chi_{kl}(r)=r\overline R_{kl}(r)$ for the free and
scattered particles respectively we find:
\begin{eqnarray}\label{phase_funct_equation}
\chi_{kl}\frac{\partial\overline{\chi}_{kl}}{\partial
r}-\overline{\chi}_{kl}\frac{\partial\chi_{kl}}{\partial
r}-\beta'\hbar^2\left(\chi_{kl}\frac{\partial^3\overline{\chi}_{kl}}{\partial
r^3}-\overline{\chi}_{kl}\frac{\partial^3\chi_{kl}}{\partial
r^3}-\right.\nonumber
\\
\left.\frac{\partial\chi_{kl}}{\partial
r}\frac{\partial^2\overline{\chi}_{kl}}{\partial
r^2}+\frac{\partial\overline{\chi}_{kl}}{\partial
r}\frac{\partial^2 {\chi}_{kl}}{\partial
r^2}\right)+2\beta'\hbar^2\frac{l(l+1)}{r^2}\times\nonumber
\\
\left(\chi_{kl}\frac{\partial\overline{\chi}_{kl}}{\partial
r}-\overline{\chi}_{kl}\frac{\partial\chi_{kl}}{\partial
r}\right)=\int^{r}_{0}\chi_{kl}(r')U\overline{\chi}_{kl}(r'){\rm
d}r'.
\end{eqnarray}
It is a very complicated problem to solve the last equation, but
most interesting for us is the phase shift at large distances from
the scattering center, so we can use asymptotic wave functions
(\ref{asympt_free}) and (\ref{asymp_scatt}). Taking into account
that the last term in the left hand side of relation
(\ref{phase_funct_equation}) goes to zero at large distances from
the scatterer we obtain:
\begin{equation}\label{phase_sh_int}
\sin{\delta_l}=\frac{-1}{k(1+2\beta'\hbar^2k^2)}\int^{r}_{0}\sin{\left(kr'-l\frac{\pi}{2}\right)}U({\bf
r'},{\bf p}')\sin{\left(kr'-l\frac{\pi}{2}+\delta_l\right)}{\rm
d}r'.
\end{equation}
In so far as we consider the phase shifts at large distances from
the scattering center we can tend to infinity the upper bound in
integral (\ref{phase_sh_int}). In the first approximation we can
put in right hand side of (\ref{phase_sh_int}) asymptotic wave
function of free particle  $\chi_{kl}(r)$ instead of
$\overline\chi_{kl}(r)$, so we have:
\begin{equation}\label{phase_infty_Born}
\sin{\delta_l}=\frac{-1}{k(1+2\beta'\hbar^2k^2)}\int^{+\infty}_{0}\sin{\left(kr'-l\frac{\pi}{2}\right)}U({\bf
r'},{\bf p}')\sin{\left(kr'-l\frac{\pi}{2}\right)}{\rm d}r'.
\end{equation}
Relation (\ref{phase_infty_Born}) corresponds the first Born
approximation in relation (\ref{scattering_amplitude}). To obtain
a more accurate estimation it is necessary to use expressions
(\ref{phase_sh_int}), but in practice enough good estimation we
obtain in the Born approximation.

\section{Conclusions}
In this paper, the elastic scattering in deformed space with
minimal length was considered. We reviewed main results obtained
earlier. One should note that differential cross-section in
deformed case equals to the square of module of the scattering
amplitude, so we have the same expression as in the ordinary
quantum mechanics. As was shown the correction to the differential
cross-section for the Coulomb potential nonanalytically depends on
deformation parameters but it tends to zero when the parameters go
to zero. This dependence is caused by the shifted expansion for
the distance operator. Then we also considered the partial wave
method in deformed space. It was shown that in deformed case
scattering on spherically symmetric potential is similar to the
ordinary one. The particles that have different angular momentum
(different orbital quantum numbers $l$) are scattered
independently. So the expression for the scattering amplitude as
sum of a partial amplitudes and the relation for cross-section
formally coincide with the corresponding relations in ordinary
quantum mechanics.

\section{Aknowlegments}
The author is grateful to Prof. V. M. Tkachuk for many useful
discussions and comments. I would also like to thank Dr. A. A.
Rovenchak  for a careful reading of the manuscript.

\end{document}